# Thermodynamics of Third Order Phase Transition: A Solution to the Euler – Lagrange Equations

**♣1 E C Ekuma; 2 G C Asomba and 2 C M I Okoye**

1. Department of Physics, Southern University and A&M College, Baton Rouge, Louisiana, 70813, USA
2. Department of Physics and Astronomy, University of Nigeria, Nsukka, Nigeria

**Abstract:** The thermodynamics expected of systems undergoing third order phase transition has been investigated by identifying the orders through the analytic continuation of the functional of the free energy, using Ehrenfest thermodynamic theory. We developed the Euler – Lagrange equations for the order parameter and the vector potential and solved them for the first time using well – known mathematical formulations.

## Introduction

The general theory of the thermodynamics in materials, especially systems undergoing phase transition(s) are generally complicated. Mean field theories have provided a way to analyze these systems, and has remained a vital tool in studying complex systems like the superconductors [1,2]. Even at these opportunities provided by mean field theories, for systems undergoing higher order phase transitions (>2), there abound many divergent views in literature to the existence and non-existence of phase transition of orders strictly greater than II [see 3,4,5]. Certainly, third order phase transition exists (even higher orders) since, for

♣ Electronic Address: panaceamee@yahoo.com

a phase transition to be a second order, the relation $\int_0^{T_c} \left[ \delta C_e \left( H,T \right) / T \right] dT = 0$ must hold. However, as observed in many superconductors (for instance $Na_xCoO_2 \cdot yH_2O$ and $BaKBiO_3$) [6,7,8,9,10,11,12]; Lattice and continuum gauge theories [13]; Chiral model [14]; Spin glasses [15]; DNA under mechanical strain [16]; Ferromagnetic and antiferromagnetic spin models with temperature driven transitions [17]; Invar types alloys [18]; and many other materials and systems, this conservation law is violated. Thus, we assert that, the non-detection of phase transitions of orders strictly greater than two might have been due to the erroneously hasty generalization that all that departs from phase transition of order two can always be explained in terms of thermodynamic fluctuation.

The aim of present study is thus, to develop the thermodynamics expected of systems undergoing third order phase transition and most importantly, to solve the highly non – linear Euler – Lagrange equations for both the order parameter and the vector field using the system parameter from the local free energy of the functional of the free energy. The solution of these Euler – Lagrange equations has not been solved before by any group to the best of our knowledge.

**Theoretical Framework**

We will adopt a theoretical approach based on the classification of order of phase transitions by Ehrenfest [11,18,19] which is in terms of the ordered state free energy near the phase boundary and the functional of the system in terms of the local free energy. The validity of using Ehrenfest classification of order of phase transition has been questioned by many authors [20,21] which has erroneously led to the classification of order of phase transitions into first order and continuous phase transitions. However, as has been shown by Hilfer [22], that rewriting the singular part of the local free energy within a restricted curve through the critical point in terms of the finite difference quotient and analytic continuation in *p* allows one to classify continuous phase transitions precisely according to their orders.

According to classification of order of phase transitions by Ehrenfest, the transition in general can be of any order [11,22]. In a second order phase transition, the specific heat and the susceptibility, which are the second derivatives of the free energy (with respect to temperature and magnetic field, respectively) are discontinuous at the transition line [23]. Often, this discontinuity is replaced by a weak (or in general logarithmic singularity). Thus, one could view the Ehrenfest definition of an order of phase transition as one where the lower derivatives of the free energy are continuous at the transition but the higher derivatives are discontinuous. For a typical third order phase transition, all second order derivatives are continuous and all fourth order derivatives are singular at the transition. The third order derivatives are either discontinuous or a weakly logarithmic singular.

The equilibrium thermodynamics is completely determined by the function, $F\left[T,\langle\psi\rangle\right]$ where $\langle\psi\rangle$ is the local order parameter [24,25]. However, $F$ must be invariant under the symmetry group say, *G* of the disordered phase in order to minimize the total energy [1]. Examples of such behavior are commonly found in magnetic domains in ferromagnetic [18] where the Ising (*N* = *1*) Hamiltonian has been applied [21]. In general, $F$ is a very complicated functional of $\langle\psi\rangle$ but since in the vicinity of critical line, $\langle\psi\rangle$ is essentially zero for $T > T_c$, we will follow William – Bragg's theory [24] and expand $F$ in a Taylor series in $\langle\psi\rangle$. To make $\langle\psi\rangle$ to be spatially uniform in equilibrium in the ordered phase, we essentially for all cases redefine the order parameter. This suggests that $F$ be expressed in terms of a local free energy density, $f\left[T,\langle\psi(x)\rangle\right]$ which is a function of the field $\langle\psi(x)\rangle$ at the point $x$ only and also a part producing the energy cost for deviations from spatial uniformity [24]. After coarse graining [1,24], in its simplest form based on lattice formulation of Ginzburg – Landau theory [1,26], $F$ is give by

$$F = \int d^d x f\left[T, \langle \psi(x) \rangle\right] + \frac{1}{2}\int d^d x c\left(\nabla \langle \psi(x) \rangle\right)^2 \tag{1}$$

where $c$ is a phenomenological constants known as the coupling constant or the gradient term and $f\left[T, \langle \psi(x) \rangle\right]$ is the local free energy density, which is well known in the $\psi^4 - theory$. It must be asserted that Eq. (1) fails in the short $\lambda$ - limit. However, we make haste to add that the short $\lambda$ - limit is never a problem as most system undergoing higher order phase transitions in the vicinity of the critical points are controlled by long - $\lambda$ fluctuations and as such, Eq.(1) is adequate for our problem so long as the short - $\lambda$ is significantly suppressed [27,28]. This is accomplished by the so – called hard cutoff mechanism where excitation of wave – number greater than a cutoff $\Lambda$□ , (where $a$ is a length of order of the same range as the inter-particle interactions) are restricted in the partition trace [1].

Eq. (1) will be generalized to suit the thermodynamic theories of higher (>2) order phase transitions that will be developed, which is totally different from those developed for second order.

Following [19,29,30] and using Eq. (1), the free energy can be generalized to Landau – like from as:

$$F_p(\psi, T) = \int d^d r |\psi|^{2(p-2)} \left\{-a_p t |\psi|^2 + b_p |\psi|^4 + c_p |\nabla \psi|^2\right\} \quad \forall p > 2 \tag{2}$$

where the integer $p$ is the order of the phase transition, $a_p = a_p^o(1 - T/T_c)$, $b$□ , the term $(\nabla \times A)^2$ has been removed for brevity and it should be noted that due to gauge transformation $\nabla \to \nabla + 2\pi i A/\phi_o$; where $\phi_o = h/2e$ is the superconducting flux quantum and $A$ is the magnitude of the vector potential. Note that $F$ is a functional and as such, the corresponding (Gibb's) free energy is to be identified with its minimum. For $p = 3$, Eq.(2) becomes:

$$F_3\left(\psi = \Delta e^{i\varphi}\right) = \int d^d r \left\{ -a_3 t|\psi|^4 + b_3|\psi|^6 + c_3\left|\left(\nabla + \frac{2\pi i}{\phi_o}A\right)\psi^2\right|^2 + \frac{1}{2\mu_o}(\nabla \times A)^2 \right\} \tag{3}$$

Eq.(3) is the functional of the local free energy that contains all the thermodynamical information of the system.

**Analysis and Solution to the Functional of the System**

A critical look at Eq.(3) shows that it contains two important parameters that describes the thermodynamics of the system: the order parameter, $\psi$ and the vector potential, $A$. We will solve the local free energy of Eq.(3) using variational principles:1) with respect to the order parameter and 2) with respect to the vector potential respectively with and without any boundary conditions.

***Case I: Neglecting Gradient Term and Absence of any Field***

In the absence of any field and neglecting the gradient (coupling) term, we can write the local free energy of Eq.(3) as:

$$f_3 = -a_3 t|\psi|^4 + b_3|\psi|^6 \tag{4}$$

which, on minimization reduces to

$$\frac{df_3}{d|\psi|} = -4a_3 t|\psi|^3 + 6b_3|\psi|^5 = 0\ ;\ \Rightarrow \quad \left.\begin{aligned} |\psi_o| &= \frac{2a_3 t}{3b_3} = \frac{2a_3^o}{3b_3}\left(1 - \frac{T}{T_c}\right); \quad T < T_c \\ &= 0 \qquad\qquad\qquad\qquad ; \quad T > T_c \end{aligned}\right\} \tag{5}$$

Using Eq.(5) in (4),

$$\left\langle f_3 \right\rangle = -a_3\left(\frac{2a_3t}{3b_3}\right)^2 + b_3\left(\frac{2a_3t}{3b_3}\right)^3 = \frac{4a_o^3}{27b_3^2}\left(1-\frac{T}{T_c}\right)^3 \qquad (6)$$

In its condensed degree of freedom, we can investigate the various thermodynamic quantities of interest using Eq.(6). The entropy, $S = \partial\left\langle F\right\rangle / \partial T$ and the specific heat, $C = -T\,\partial^2\left\langle F\right\rangle / \partial T^2$ are respectively

$$S_3 = \frac{\partial\left\langle F\right\rangle}{\partial T} = -\frac{4}{9}\frac{a_o^3}{b_3^2T_c}\left(1-\frac{T}{T_c}\right)^2 \quad \text{and} \quad C_3 = \frac{8}{9}\frac{a_o^3T}{b_3^2T_c^2}\left(1-\frac{T}{T_c}\right) \qquad (7)$$

The order parameter, $\psi_o$ can be expressed in terms of the local free energy (see Eq.(6)) as

$$f\left(T\right) = \frac{b_3}{2}\psi_o^6; \quad \psi_o^2 = 2a_3t/3b_3; \; t>0 \; (t=0; \; t<0) \qquad (8)$$

***Case II*: *Presence of Gradient Term and Field***

Next, we investigate, the effect of presence of filed and the incorporation of the gradient term to the thermodynamic behavior of the system. Again, with recourse to Eq.(3), we apply variational principles first with respect to the order parameter and secondly with respect to the potential field.

In terms of the order parameter, we obtain that

$$c_3\nabla^2\psi + c_3\frac{\left|\nabla\psi\right|^2}{\psi} + 2a_3t\psi - 3b_3\psi^3 = 0 \qquad (9)$$

which is the Euler – Lagrange equation of the free energy for the order parameter: $\psi = \Delta e^{i\varphi}$ in the third order phase transition regime. Also, for the vector potential, $A$, we have,

$$\frac{1}{\mu_o \lambda^2} A = \frac{e^*}{m^*}\left( i\hbar \qquad \frac{e^*}{c^{-}} \qquad \right) |\psi|^4 \tag{10}$$

where $e^*$ and $m^*$ are the effective charge and effective mass respectively. In the Meissner state at the depth $\lambda$, the gradient term can be neglected to obtain the Euler – Lagrange equation for the vector potential in terms of the penetration length, $\lambda$ as:

$$\lambda^{-2} = \frac{e^{*2}}{c^2 m^*} \mu_o |\psi|^4 = 2\mu_o c \left( \frac{2\pi}{\phi_o} \right)^2 |\psi|^4 \tag{11}$$

which, is the Euler – Lagrange equation in terms of the penetration length; and can be rewritten using gauge transformation as,

$$\nabla^2 A - 2\mu_o c \left( \frac{2\pi}{\phi_o} \right)^2 |\psi|^4 A = 0 \tag{12}$$

which is nothing but the London equation for the behavior of the system in third order phase transition regime.

The specific heat can be rewritten in terms of the penetration length by using Eq.(11) in Eq.(8) as

$$C_3(T) = \frac{b_3 T}{2(2\mu_o c)^{3/2}} \left( \frac{\phi_o}{2\pi} \right)^3 \frac{\partial^3}{\partial T^3} \lambda^{-3} \tag{13}$$

***Solution to the Euler – Lagrange Equations***

We will seek the solutions to Eqs.(9 &12) which are highly non – linear Euler – Lagrange equations for the order parameter and the vector potential respectively.

As a starting point, Eq.(12) can be redefine in terms of the potential, $\Omega(r)$ as

$$\nabla^2 A + \Omega(r) A = 0 \tag{14}$$

where $\Omega = -2\mu_0 c\left(\frac{2\pi}{\phi_0}\right)^2 |\psi_0|^4 = -\lambda^{-2}$ and $\lambda(r)^{-2} = 2\mu_0 c\left(\frac{2\pi}{\phi_0}\right)^2 \left[\frac{1-\frac{r}{32}}{2b}\right]$

$$\therefore \quad \Omega(r) = -\lambda(r)^{-2} = 2\mu_0 c\left(\frac{2\pi}{\phi_0}\right)^2 \left[\frac{\frac{r}{32}-1}{2b}\right] \tag{15}$$

Recall that $\nabla^2 = \frac{\partial^2}{\partial x^2} + \frac{\partial^2}{\partial y^2} + \frac{\partial^2}{\partial z^2}$, and the function $\Omega(r)$ will be regarded in this case as a potential function independent of $\theta$ and $\phi$ when we transform the Cartesian coordinates $(x, y, z)$ to polar $(r, \theta, \phi)$. Thus, Eq.(14) reduces to

$$\frac{1}{r^2}\left[\frac{\partial}{\partial r}\left(r^2 \frac{\partial A}{\partial r}\right) + \frac{1}{\sin\theta}\frac{\partial}{\partial\theta}\left(\sin\theta \frac{\partial A}{\partial\theta}\right) + \frac{1}{\sin^2\theta}\frac{\partial^2 A}{\partial\phi^2}\right] + \Omega(r) A = 0 \tag{16}$$

which on using the method of separation of variables: $A = R(r) S(\theta, \phi)$, gives

$$\frac{1}{R}\left[\frac{d}{dr}\left(r^2 \frac{dR}{dr}\right) + \Omega(r) r^2\right] = -\frac{1}{S}\left[\frac{1}{\sin\theta}\frac{\partial}{\partial\theta}\left(\sin\theta \frac{\partial S}{\partial\theta}\right) + \frac{1}{\sin^2\theta}\frac{\partial^2 S}{\partial\phi^2}\right] \tag{17}$$

Since the LHS is depends only on $r$ and RHS is independent of $r$, therefore each side must be equal to the same constant say, $\mu^2$ for the equation to be valid. Thus, from Eq.(17), we have for the LHS that;

$$\frac{1}{R}\left[\frac{d}{dr}\left(r^2 \frac{dR}{dr}\right) + \Omega(r) r^2\right] = \mu^2 \Rightarrow \qquad \frac{1}{r^2}\frac{d}{dr}\left(r^2 \frac{dR}{dr}\right) + \left[\Omega(r) - \frac{\mu^2}{r^2}\right] R = 0 \tag{18}$$

Provided $\Omega(r)=\lambda^{-2}$, Eq.(18) reduces to Bessel's equation, with general solutions;

$$R(r)=A_1 J_\mu\left(\lambda^{-2}r\right)+B_1 J_{-\mu}\left(\lambda^{-2}r\right) \text{ for fractional } \mu \tag{19a}$$

$$R(r)=A_1 J_\mu\left(\lambda^{-2}r\right)+B_1 Y_\mu\left(\lambda^{-2}r\right) \text{ for integral } \mu \tag{19b}$$

Similarly, taking the RHS of Eq.(17),

$$-\frac{1}{S}\left[\frac{1}{\sin\theta}\frac{\partial}{\partial\theta}\left(\sin\theta\frac{\partial S}{\partial\theta}\right)+\frac{1}{\sin^2\theta}\frac{\partial^2 S}{\partial\phi^2}\right]=\mu^2 \Rightarrow \frac{1}{\sin\theta}\frac{\partial}{\partial\theta}\left(\sin\theta\frac{\partial S}{\partial\theta}\right)+\frac{1}{\sin^2\theta}\frac{\partial^2 S}{\partial\phi^2}=-\mu^2 S \tag{20}$$

Again, applying separable, $S(\theta,\phi)=\Theta(\theta)\Phi(\phi)$, Eq.(20) reduces to

$$\frac{1}{\Theta}\left[\sin\theta\frac{d}{d\theta}\left(\sin\theta\frac{d\Theta}{d\theta}\right)\right]+\mu^2\sin^2\theta=-\frac{1}{\Phi}\frac{d^2\Phi}{d\phi^2}=l^2 \tag{21}$$

where $l$ is constant of separation. Taking the RHS of Eq.(21), we have;

$$-\frac{1}{\Phi}\frac{d^2\Phi}{d\phi^2}=l^2 \Rightarrow \frac{d^2\Phi}{d\phi^2}+l^2\Phi=0 \Rightarrow \left(D^2+l^2\right)\Phi=0 \Rightarrow \quad D=\pm il;\ \ \Phi\neq 0 \tag{22}$$

with the solution

$$\Phi(\phi)=e^{il\phi}=\cos l\phi+i\sin l\phi. \tag{23}$$

For the LHS of Eq.(21), we have;

$$\frac{1}{\Theta}\left[\sin\theta\frac{d}{d\theta}\left(\sin\theta\frac{d\Theta}{d\theta}\right)\right]+K\sin^2\theta=l^2 \Rightarrow \frac{1}{\sin\theta}\frac{d}{d\theta}\left(\sin\theta\frac{d\Theta}{d\theta}\right)+\left\{n(n+1)-\frac{l^2}{\sin^2\theta}\right\}\Theta=0 \tag{24}$$

For convenience we have set $\mu^2=n(n+1)$. Making change of variable; i.e. $\cos\theta=u$ so that $\frac{1}{\sin\theta}\frac{d}{d\theta}\equiv-\frac{d}{du}$, then Eq.(24) reduces to;

$$\frac{d}{du}\left\{\left(1-u^2\right)\frac{d\Theta}{du}\right\}+\left\{n\left(n+1\right)-\frac{l^2}{1-u^2}\right\}\Theta=0 \tag{25}$$

Eq.(25) is the associated Legendre equation, and has its solution as;

$$\Theta = AP_n^m\left(u\right)+BQ_n^m\left(u\right)=AP_n^m\left(\cos\theta\right)+BQ_n^m\left(\cos\theta\right) \tag{26}$$

where $A$ and $B$ are constants to be determined by the initial conditions. Hence, we obtain the complete general solution to the Euler – Lagrange equation for the vector potential as

$$A\left(r,\theta,\phi\right)=\left(A_1J_\mu\left(\lambda r\right)+B_1J_{-\mu}\left(\lambda r\right)\right)\left(\cos l\phi+i\sin l\phi\right)\left(A_2P_n^m\left(\cos\theta\right)+B_2Q_n^m\left(\cos\theta\right)\right) \tag{27}$$

For the order parameter, since *c* is a gradient term, we can rewrite Eq.(9) as:

$$\nabla^2\psi+\frac{\left|\nabla\psi\right|^2}{\psi}+\frac{2t\psi-3b\psi^3}{c}=0 \tag{28}$$

And recalling that $\left|\nabla\psi\right|=\left|\left(\frac{\partial\psi}{\partial x},\frac{\partial\psi}{\partial y},\frac{\partial\psi}{\partial z}\right)\right|=\sqrt{\left(\frac{\partial\psi}{\partial x}\right)^2+\left(\frac{\partial\psi}{\partial y}\right)^2+\left(\frac{\partial\psi}{\partial z}\right)^2}$, we have that,

$\left|\nabla\psi\right|^2=\left(\frac{\partial\psi}{\partial x}\right)^2+\left(\frac{\partial\psi}{\partial y}\right)^2+\left(\frac{\partial\psi}{\partial z}\right)^2$. Solving in 1D,

$$\frac{\partial^2\psi}{\partial x^2}+\frac{1}{\psi}\left(\frac{\partial\psi}{\partial x}\right)^2+\frac{2t\psi-3b\psi^3}{c}=0,\quad \psi=\psi\left(t,x\right) \tag{29}$$

Assuming a constant temperature gradient field (such that *t* is a constant) where the variation of the system depends entirely on the order parameter, further assuming natural units, *c = 1* and also noting that *b >1* and as such also a constant, we can rewrite Eq.(29) as:

$$\frac{d^2\psi}{dx^2}+\frac{1}{\psi}\left(\frac{d\psi}{dx}\right)^2+2\psi-3\psi^3=0 \tag{30}$$

To linearizing the non – linear differential (Eq.(30)) we use a numerical procedure by letting $z_1=\psi$, thus reducing to,

$$z_1''+\frac{1}{z_1}\left(z_1'\right)^2+2z_1-3z_1{}^3=0 \tag{31}$$

Next, let let $z_2=z_1'$ hence we have

$$z_2'+\frac{1}{z_1}z_2^2+2z_1-3z_1{}^3=0 \qquad \Rightarrow \qquad z_2'=-\frac{1}{z_1}z_2^2-2z_1+3z_1{}^3 \tag{32}$$

$$\text{where } \begin{cases} z_1'=z_2 \\ z_2'=-\dfrac{1}{z_1}z_2^2-2z_1+3z_1{}^3 \end{cases} \tag{33}$$

Observe that Eq.(28) has been reduced to a system of first order differential equations which can then be solved.

**Conclusion**

The thermodynamics of third order phase transition have been analyzed using mean field theory based on Ehrenfest thermodynamics. Certain thermodynamical relations well – known for systems undergoing second order phase transitions are recovered for third order phase transition. We developed equations for the phase boundary in terms of discontinuities in thermodynamic observables, derived Ginzburg – Landau free energies, developed Euler – Lagrange equations for the order parameter and the vector potential, and solved the highly non – linear Euler – Lagrange equations for both the order parameter and the vector potential.

**Acknowledgement**

This work was supported by the Government of Ebonyi State, Federal Republic, Nigeria through its superior award for overseas training to one of the authors ECE. The authors wish to acknowledge insightful discussions from Prof. Diola Bagayoko, Prof. Ali Fazley, and Dr. Maliki.